# Ab-initio Study of Structural, Magnetic, Optoelectronic and Thermo-Physical Properties of HoPdBi Half-Heusler Semimetal

Tanvir Khan, F. Parvin, S. H. Naqib*

Department of Physics, University of Rajshahi, Rajshahi 6205, Bangladesh

*Corresponding author email: salehnaqib@yahoo.com

**Abstract**

In this investigation, we have used the density functional theory (DFT) to investigate several aspects of the half-Heusler compound HoPdBi. The following properties have been studied: spin polarized electronic properties, magnetic moment, phonon dispersion with phonon density of states, structural, elastic properties, optical characteristics, and thermo-physical features. The calculated unit cell volume and ground-state lattice characteristics closely match the experimental results. This study is the first to examine the optoelectronic, thermo-physical, and elastic characteristics of HoPdBi. The mechanical stability requirements were met by the calculated elastic constants. The compound's ductility is shown by the estimated Pugh's ratio, Poisson's ratio, and Cauchy pressure. Band structures and electronic energy density of states have been evaluated in order to better understand the magnetic features with spin polarization. Band structure simulations were conducted with and without the spin-orbit coupling (SOC) effect in order to look into any topological signature. The electrical band structure of the compound shows semi-metallic properties. The reflectivity, absorption coefficient, refractive index, dielectric function, optical conductivity, and loss function of this semi-metal have all been thoroughly examined. The compound is a good reflector in infrared region and a good absorber of ultraviolet (UV) light. This compound is a suitable candidate for high temperature applications and possesses potential as heat sink because of its high melting point and thermal conductivity. It is also suitable for spintronics applications. The majority of this study's findings are completely novel.

## 1. Introduction

Fritz Heusler's discovery in 1903 that an alloy with the composition $Cu_2MnAl$ acts like a ferromagnet despite the fact that none of its constituent elements are magnetic on their own [1] marks the beginning of one of the most fascinating material classes. Heusler compounds are the current name for this extraordinary substance and its counterparts, which together make up a huge collection of around 1000 compounds. They are metallic or ternary semiconducting materials with a 2:1:1 or 1:1:1 stoichiometry (sometimes referred to as "Half-Heusler"). Remarkably, merely counting the amount of valence electrons can predict the characteristics of many Heusler compounds [2]. Superconducting Heusler compounds, for instance, have about 27 valence electrons and are not magnetic. With almost 250 members, semiconductors exhibit yet another significant subclass and are regarded as innovative materials for energy systems. It is simple to adjust their band gaps from 0 to 4 eV by altering their chemical makeup. They so garnered a lot of interest as viable options for thermoelectric and solar cell applications. Indeed, TiNiSn-based materials have recently been shown to have outstanding thermoelectric characteristics [3]. Multifunctional topological insulators, a new state of matter in which spin-polarized edge and surface states are topologically protected against impurity scattering, are a new class of Heusler compounds that were recently predicted based on their computed electronic band structures [4,5]. Ternary compounds make it simple to introduce multifunctionality, or the combining of two or more functionalities, including superconductivity and topological edge states, in a single material [5]. The broad category of magnetic $X_2YZ$ compounds exhibit a wide range of magnetic behavior and multifunctional magnetic features, including magneto-structural [6], magneto-optical [1], and magnetocaloric [7] characteristics. For electrons with one spin orientation, the half-metallic ferromagnets, which belong to the broad family of magneto-electrical Heusler compounds, are semiconducting; for electrons with the opposite spin orientation, they are metallic. These substances are suitable materials for spintronic applications because they have conduction electrons that are almost completely spin polarized.

Half-Heusler materials, usually XYZ, can be thought of as compounds with both an ionic and a covalent component. While Z can be thought of as the anionic counterpart, X and Y atoms have a definite cationic nature. All three of the possible permutations can be found because the nomenclature used in literature differs greatly and can be arranged alphabetically, based on electronegativity, or randomly. The formula starts with the element that is most electropositive. It may be a transition metal, a rare earth element, or an element of the main group. A major group element from the second half of the periodic table, such as Li, Al, Si, Zr, Ni, Sn are the electronegative elements at the end [1,8].

We have selected this half-Heusler HoPdBi because of its intriguing characteristics and potential for applications in a range of structural and functional fields. In half-Heusler semimetal (HoPdBi), antiferromagnetism and superconductivity ($T_c = 0.7K$) have already been investigated [9]. Nevertheless, many important properties and traits of HoPdBi are not known currently. Most of them are crucial for prospective applications; for example, optoelectronic properties, thermo-physical aspects, and elastic behavior have not yet been fully investigated. The examination of Cauchy pressure, Pugh's ratio, Poisson's ratio, micro- and macro-hardness, machinability index, elastic moduli, and many other bonding-related aspects still require investigation. The Debye temperature, melting temperature, lattice thermal conductivity, sound velocities, dominating phonon wavelength, and lowest thermal conductivity are among the thermo-physical parameters that have not yet been investigated thoroughly. The behavior of a material at different temperatures can be inferred from these thermo-physical characteristics. In this work, we have examined these attributes. Furthermore, understanding optical properties is essential when selecting a system for use in optoelectronic devices. Thus, for the selected compound, optical parameters have been investigated.

This work's remaining sections are organized as follows: A synopsis of the computational methodology employed in this investigation is provided in section 2. We present the computational results and analyses in section 3. The main characteristics and pertinent findings for this work are covered in section 4.

## 2. Computational methodology

The density functional theory (DFT), which solves the Kohn-Sham equation [10] with periodic boundary conditions (including Bloch states), is the most widely used framework for ab-initio calculations on crystalline materials. The DFT-based CAmbridge Serial Total Energy Package (CASTEP) was used to examine the compound’s physical properties [11]. The total energy plane-wave pseudopotential is a method used in this code. Generalized gradient approximations (GGA) of the Perdew-Burke-Ernzerhof (PBE) [12], GGA of the Perdew-W91 [13], GGA of the RPBE [14], GGA of the PBEsol [12], and local density approximation (LDA) [15,16] were all used in the present investigation. The importance of accurate geometry optimization is paramount. LDA of the CA-PZ yields the best structural properties for the HoPdBi. Ultra-soft Vanderbilt type pseudopotentials were employed to compute the electron-ion interactions [17]. This method significantly reduces calculation time without compromising accuracy by loosening the norm-conserving requirement and producing a smooth and computationally friendly pseudopotential. This study uses the Monkhorst-Pack technique [18] of size 10 × 10 × 10 for the first irreducible Brillouin zone (BZ) k-point sampling. The cut-off energy for the plane wave expansion was selected to be 550 eV. For further details on pseudopotentials and k-point grids, we refer to [19]. The *stress-strain* method was used to determine the cubic structure's single crystal elastic constants ($C_{ij}$) [20]. All of the elastic characteristics, such as the bulk modulus (B), shear modulus (G), and Young's modulus, can be calculated using the Voigt-Reuss-Hill (VRH) formalism [21,22] using the single crystal elastic constants $C_{ij}$. Electronic band structure, partial density of states (PDOS), and total density of states (TDOS) are computed using the results of the optimized geometry of HoPdBi. It is possible to obtain all the optical constants that depend on the energy or frequency of the electromagnetic wave using the complex dielectric function, $\varepsilon(\omega) = \varepsilon_1(\omega) + \varepsilon_2 i(\omega)$. The real part of the dielectric function is computed using the imaginary part via Kramers-Kronig equation [23]. The imaginary part, $\varepsilon_2(\omega)$, of the complex dielectric function can be computed using the momentum representation of the matrix elements of the photon-induced transition between occupied and unoccupied electronic states. The CASTEP-contained expression for the dielectric function's imaginary component is given by:

$$\varepsilon_2(\omega) = \frac{2e^2\pi}{\Omega\varepsilon_0}\sum_{kvc}|< \Psi_k^c|\hat{u}.\vec{r}|\Psi_k^v > |^2\,\delta(E_k^c - E_K^v - E) \quad (1)$$

The unit cell's volume is denoted by $\Omega$, the incident electromagnetic wave (photon) angular frequency (or energy) by ω, the unit vector defining the incident electric field's polarization direction by $\hat{u}$, the electronic charge by e, and the conduction and valence band wave functions at a given wave-vector k by $\Psi_k^c$ and $\Psi_k^v$ , respectively. The delta function in Equation (1) implements the conservation of momentum and energy throughout the optical transition. Using the interrelations [24,25] below, it is possible to infer all other important optical features from the dielectric function ε(ω) once it is known at different energies. The real, n(ω), and imaginary, k(ω), components of the complex refractive index can be calculated using the following relations:

$$n(\omega) = \frac{1}{\sqrt{2}}[\{\varepsilon_1(\omega)^2 + \varepsilon_2(\omega)^2\}^{\frac{1}{2}} + \varepsilon_1(\omega)]^{1/2} \tag{2}$$

$$k(\omega) = \frac{1}{\sqrt{2}}[\{\varepsilon_1(\omega)^2 + \varepsilon_2(\omega)^2\}^{\frac{1}{2}} - \varepsilon_1(\omega)]^{1/2} \tag{3}$$

Again, using the complex refractive index components, the reflectivity, R(ω), may be calculated:

$$R(\omega) = \frac{(n-1)^2+k^2}{(n+1)^2+k^2} \tag{4}$$

Furthermore, the following equations can be used to calculate the absorption coefficient, α(ω), optical conductivity, σ(ω), and energy loss function, L(ω):

$$\alpha(\omega) = \frac{4\pi k(\omega)}{\lambda} \tag{5}$$

$$\sigma(\omega) = \frac{2W_{cv}\hbar\omega}{\vec{E}_0^2} \tag{6}$$

$$L(\omega) = \mathrm{Im}(-\frac{1}{\varepsilon(\omega)}) \tag{7}$$

In Equation (6), $W_{cv}$ is the optical transition probability per unit time.

The acoustic velocities in the crystal have been used to compute the Debye temperature. The remaining thermo-physical characteristics are calculated using the crystal density, elastic constants, and elastic moduli of HoPdBi. The aforementioned formalisms have been applied with excellent precision to numerous condensed phases in the past, including the technologically significant MAB and MAX phases [26-30]. For SOC, magnetic moment and spin polarized calculations, Quantum Espresso and Wien2k packages have been used.

## 3. Results and Discussion

### 3.1 Structural properties

Many physical properties, including elastic constants, electronic band structure, and optical properties, are strongly influenced by a substance's symmetry and crystal structure. The arrangement of atoms, their separations from one another, and their electronic states determine every facet of a solid's physical characteristics. HoPdBi has cubic structure $[C1_b]$ with space group F$\bar{4}$3m (no.216) [31]. Figure 1 below shows the HoPdBi compound's crystal structure. Four Ho, four Pd, and four Bi are the four formula units that make up a single unit cell of HoPdBi. Ho atoms are located at (0, 0, 1/2), Pd atoms are located at (3/4, 1/4, 3/4), and Bi atoms are located at (0, 0, 0).

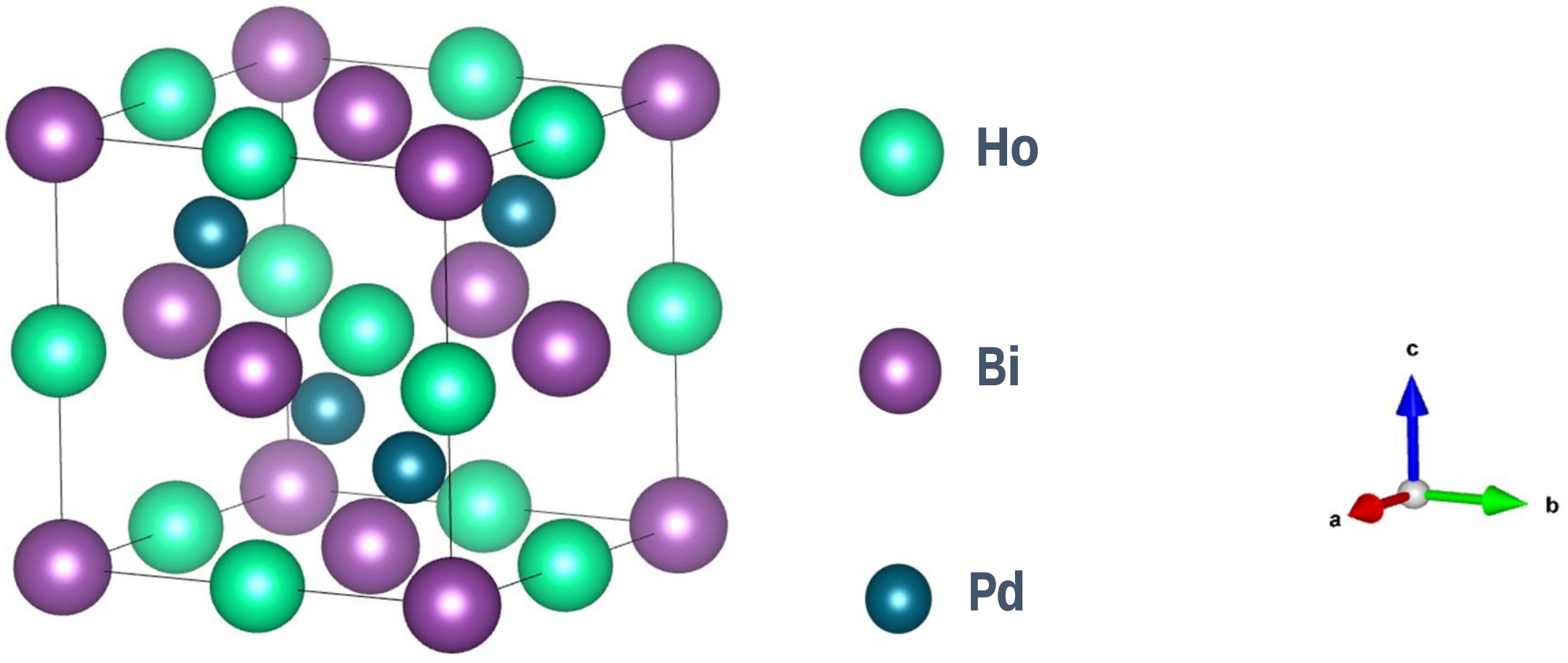

**Figure 1:** Schematic crystal structure of HoPdBi compound. The crystallographic directions are also shown.

The lattice parameters of the relevant compound are totally relaxed during the geometry optimization. The angles of the lattice are: α = β = γ = 90◦. Table 1 compares the lattice parameters derived through geometry optimization with experimental data.

**Table 1:** The optimized lattice parameters a (Å), and optimized cell volume V ( $Å^3$) of HoPdBi compound.

| Compound | a | V | Functionals | Ref. |
|---|---|---|---|---|
| HoPdBi | 6.660 | 295.42 | ---- | [31] |
| | 6.659 | 295.31 | LDA | This work |
| | 6.799 | 314.29 | GGA-PBE | This work |

### 3.2 Elastic properties

Elastic properties of solids are strongly linked with cohesive energy and atomic bonding. The mechanical behavior of solids under applied load is determined by elastic constants and moduli. The physical characteristics known as the elastic constants are used to determine the kind of interatomic bonding that exists in a solid and to gauge its mechanical strength and stability. The mechanical properties of a material, such as stability, stiffness, brittleness, ductility, and elastic anisotropy, are controlled by the elastic constants. Understanding these properties is essential when selecting a material for engineering purposes. A cubic crystal's stiffness matrix includes only three distinct elastic constants: $C_{11}$, $C_{12}$, and $C_{44}$. The Born stability criteria [32] are inequality-based conditions that guarantee the mechanical stability of crystalline structures under static stress. Different crystal classes have different stability requirements. The Born stability condition, in the case of cubic crystals, simplify to the following form [33]:

$$C_{11} > 0,\ C_{44} > 0,\ (C_{11} - C_{12}) > 0,\ (C_{11} + 2C_{12}) > 0 \qquad (8)$$

All of these requirements are met by the values of $C_{ij}$, guaranteeing the elastic stability of HoPdBi compounds.

Young's modulus (Y), Poisson's (σ) ratio, shear modulus (G), and bulk modulus (B) can all be calculated with the Voigt-Reuss-Hill (VRH) approximation. The Voigt and Reuss approximations are used to calculate the upper and lower bounds of the elastic moduli of polycrystalline materials, respectively. Between the Voigt and Reuss limits lies the actual value. A solid's resistance to compression brought on by hydrostatic pressure is indicated by its bulk modulus (B) [34].

Conversely, a solid's ability to tolerate external stress that tends to alter its shape is demonstrated by its shear modulus (G) [32]. A solid's resistance to tensile stress that alters its length is gauged by its Young's modulus (Y) [35]. This characteristic also determines a solid's rigidity. Table 2 displays the calculated values. Table 2 shows that B > G, indicating that the shearing stress of HoPdBi should regulate its mechanical stability.

The bulk modulus (B), shear modulus (G) [using the Voigt-Reuss-Hill (VRH) method], and Young's modulus (Y) were calculated using the following formulae [36-38]:

$$B_H = \frac{B_V + B_R}{2} \quad (9)$$

$$G_H = \frac{G_V + G_R}{2} \quad (10)$$

$$Y = \frac{9BG}{(3B+G)} \quad (11)$$

Generally speaking, a solid can be classified as either brittle or ductile. Three crucial factors that establish a system's brittle and ductile characteristics are the Pugh's ratio (B/G), Poisson's ratio (σ), and Cauchy pressure (C″) [39,40]. To differentiate between the brittleness and ductility of materials, Pugh put forth an empirical relationship. The ratio of the bulk and shear moduli (B/G), establishes whether a material is brittle or ductile. It is commonly recognized that a material will exhibit brittle behavior if its Pugh's ratio is less than 1.75 and ductile behavior if it is greater than 1.75 [41]. Given that HoPdBi's (B/G) value is 2.41, the compounds should be considered as ductile.

A measure of how much a material has changed form (either expanded or contracted) in relation to the direction in which it was loaded is called the Poisson's ratio (σ). This metric is also used to measure the stability of solids against shear. The range of values for a solid's Poisson's ratio is as follows [38]: $-1.0 \leq \sigma \leq 0.5$. Elastic deformation has no effect on volume if $\sigma = 0.5$ [42]. With a critical value of 0.26, this characteristic can predict if a material will be brittle or ductile. The material is brittle (ductile) if σ is lesser (higher) than 0.26 [43,44]. From table 2 we can conclude that HoPdBi is ductile. This is in line with the outcome of the Pugh's ratio. Better plasticity results from a larger Poisson's ratio. The following expression is used to calculate the Poisson's ratio:

$$\sigma = \frac{(3B-2G)}{2(3B+G)} \quad (12)$$

The nature of interatomic forces in solids can be studied by looking at the value of σ [43,45]. If this ratio remains in the range of 0.25 to 0.50, central force interaction will be dominant. Otherwise, the non-central force will assume command. Therefore, a central force should control the atoms' bonding in HoPdBi. Cauchy pressure ($C$") is another interesting mechanical characteristic. The formula for a material's Cauchy pressure is $C$" = ($C_{12}$ - $C_{44}$). Brittle materials are characterized by negative Cauchy pressure, whereas ductile materials are characterized by positive Cauchy pressure [45]. The angular properties linked to atomic bonding in a material are further elucidated by the Cauchy pressure [46]. Positive and negative Cauchy pressure are associated with the presence of ionic and covalent bonding in a material, respectively. Pettifor's rule [46] states that a material with strong ductility and a large positive Cauchy pressure has significant metallic bonding. On the other hand, a material with negative Cauchy pressure exhibits significant covalent bonding and is more brittle due to its higher number of angular bonds. Consequently, a positive Cauchy pressure value suggests that HoPdBi has metallic bonding and is ductile in nature. Table 3 provides the tetragonal shear modulus, compressibility, and all the independent elastic constants. Cauchy pressure and machinability index also has an impact on the solids' dry lubricating properties and flexibility [47-49]. Therefore, strong bonding strength and low shear resistance are combined to create good machinability and increased dry lubricity. High plastic strain values, low feed forces, good lubrication, and minimal friction are characteristics of materials with a high machinability. The following formula has been used to determine the machinability index ($\mu_M$):

$$\mu_M = \frac{B}{C_{44}} \quad (13)$$

One characteristic of solids that shows how resistant they are to localized plastic deformation is their hardness. From the perspective of the application, the material's hardness is significant. To understand how a material will respond mechanically and structurally under high stress, one must know its hardness. $C_{44}$ and G are considered to be the best markers of hardness [50,51]. Numerous formulae can be used to determine a solid's hardness. Some of the commonly used expressions are listed below:

$$H_{micro} = \frac{(1-2\sigma)}{6(1+\sigma)} \quad (14)$$

$$H_{macro} = 2[(G/B)^2G]^{0.585} \quad (15)$$

$$(H_V)_{Tian} = 0.92(G/B)^{1.137}G^{0.708} \quad (16)$$

$$(H_V)_{Teter} = 0.151G \quad (18)$$

$$(H_V)_{Mazhnik} = Y_0X(\sigma)Y \quad (19)$$

In equation (19), $X(\sigma)$ is a function of Poisson's ratio and can be written as:

$$X(\sigma) = \frac{1 - 8.5\sigma + 19.5\sigma^2}{1 - 7.5\sigma + 12.2\sigma^2 + 19.6\sigma^3}$$

$Y_0$ is a dimensionless constant with a value of 0.096. The computed values of the hardness are displayed in Table 4.

**Table 2:** Calculated bulk modulus B (in GPa), shear moduli G (in GPa), Young modulus Y (in GPa), Pugh's ratio (B/G), Poisson's ratio (σ), Cauchy pressure ($C''$) (in GPa), and machinability index ($\mu_M$) for HoPdBi compound.

| Compound | B | G | Y | B/G | σ | $C''$ | $\mu_M$ | Ref. |
|---|---|---|---|---|---|---|---|---|
| HoPdBi | 76.15 | 31.62 | 83.38 | 2.41 | 0.31 | 23.44 | 2.41 | This work |

**Table 3:** Calculated elastic constants, $C_{ij}$ (GPa), tetragonal shear modulus, $C'$ (GPa) and compressibility $\beta$ (TP$a^{-1}$) of the compound HoPdBi.

| Compound | $C_{11}$ | $C_{12}$ | $C_{44}$ | $(C_{11} - C_{12})$ | $C'$ | $\beta$ | Ref. |
|---|---|---|---|---|---|---|---|
| HoPdBi | 118.45 | 54.99 | 31.54 | 63.46 | 31.73 | 4.37 | This work |

**Table 4:** Calculated hardness (GPa) based on elastic moduli and Poisson's ratio of HoPdBi compound.

| Compound | $H_{micro}$ | $H_{macro}$ | $(H_V)_{Tian}$ | $(H_V)_{Teter}$ | $(H_V)_{Mazhnik}$ | Ref. |
|---|---|---|---|---|---|---|
| HoPdBi | 4.00 | 2.39 | 3.90 | 4.77 | 4.44 | This work |

Using the ELATE [52] code, we have shown the variation of Young's modulus (Y), shear modulus (G), Poisson's ratio (σ) and compressibility along different directions within the crystal. These plots clearly demonstrate the elastic anisotropy of HoPdBi. For an isotropic crystal, the 3D plots must be spherical; otherwise, they show the degree of anisotropy. We have shown ELATE [52] generated 3D plots of directional independences of the Young's modulus, shear modulus and Poisson's ratio for HoPdBi compound below. From Figure 2, it is clear that the 3D figures of Y, G, σ and $\beta$ don't show any departure from spherical shape, indicating isotropy. This follows from the cubic structure of the crystal under study.

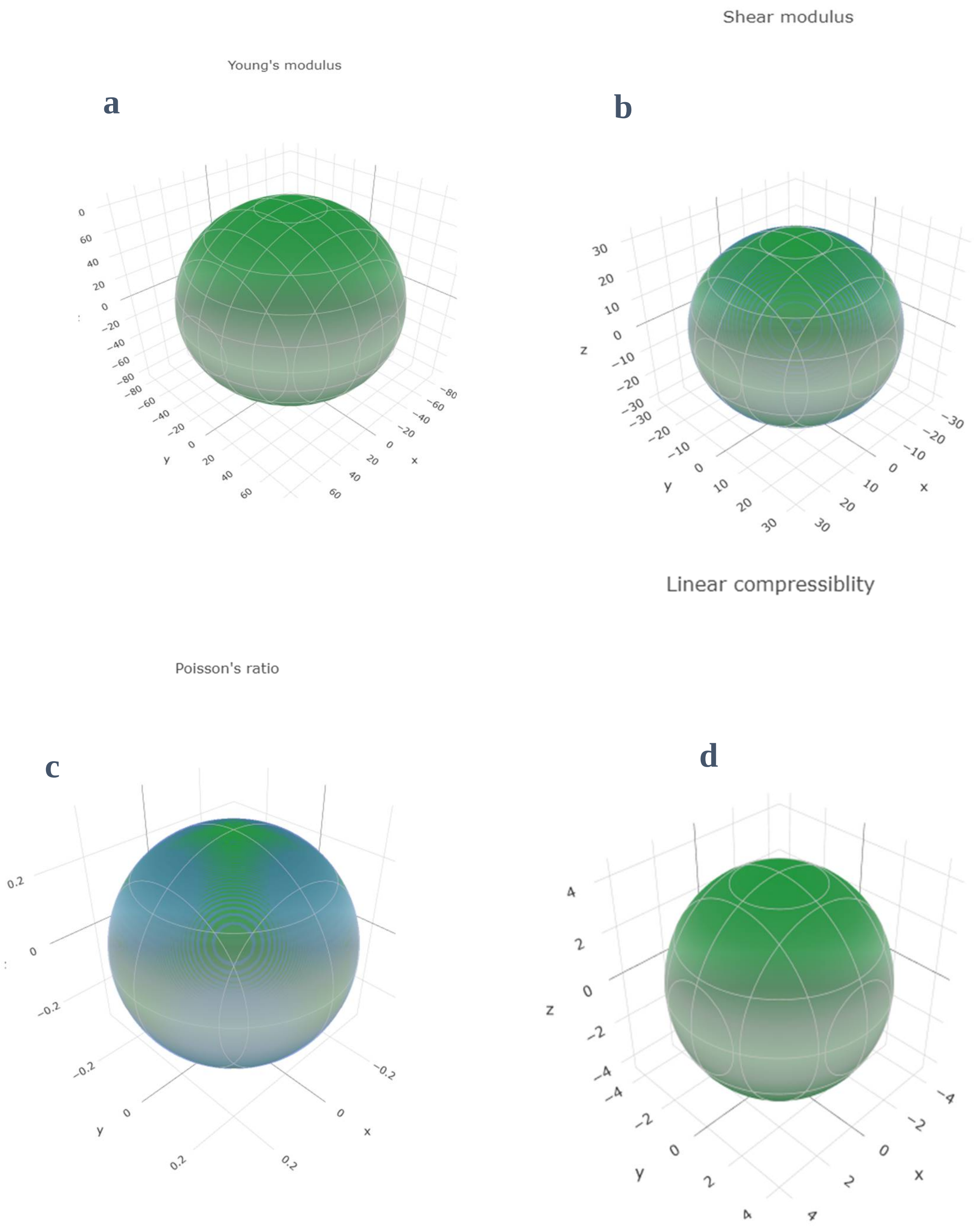

**Figure 2**: 3D directional independences in (a)Young’s modulus (Y) (b shear modulus (G) (c) Poisson’s ratio (σ) (d) linear compressibility of HoPdBi compound.

### 3.3 Electronic band structure and density of states

A solid's optical and charge transport properties are determined by its electronic band structure. Information on bonding and electronic stability can also be found from the band structure. Band topologies and charge carrier effective masses, which can be computed from the electronic band structure, are crucial for modeling nanostructures and electronic devices [53,54]. We computed the electronic band structure for the HoPdBi compound in the k-space along high symmetry directions. Figure 3 shows the spin orbit coupling effect on the electronic band structure for the HoPdBi using Quantum Espresso (QE) package. The horizontal dashed line represents the Fermi level $E_F$. The band structure indicates the semi metallic nature of HoPdBi compound as at the Fermi level, conduction bands and valance bands are slightly overlapping.

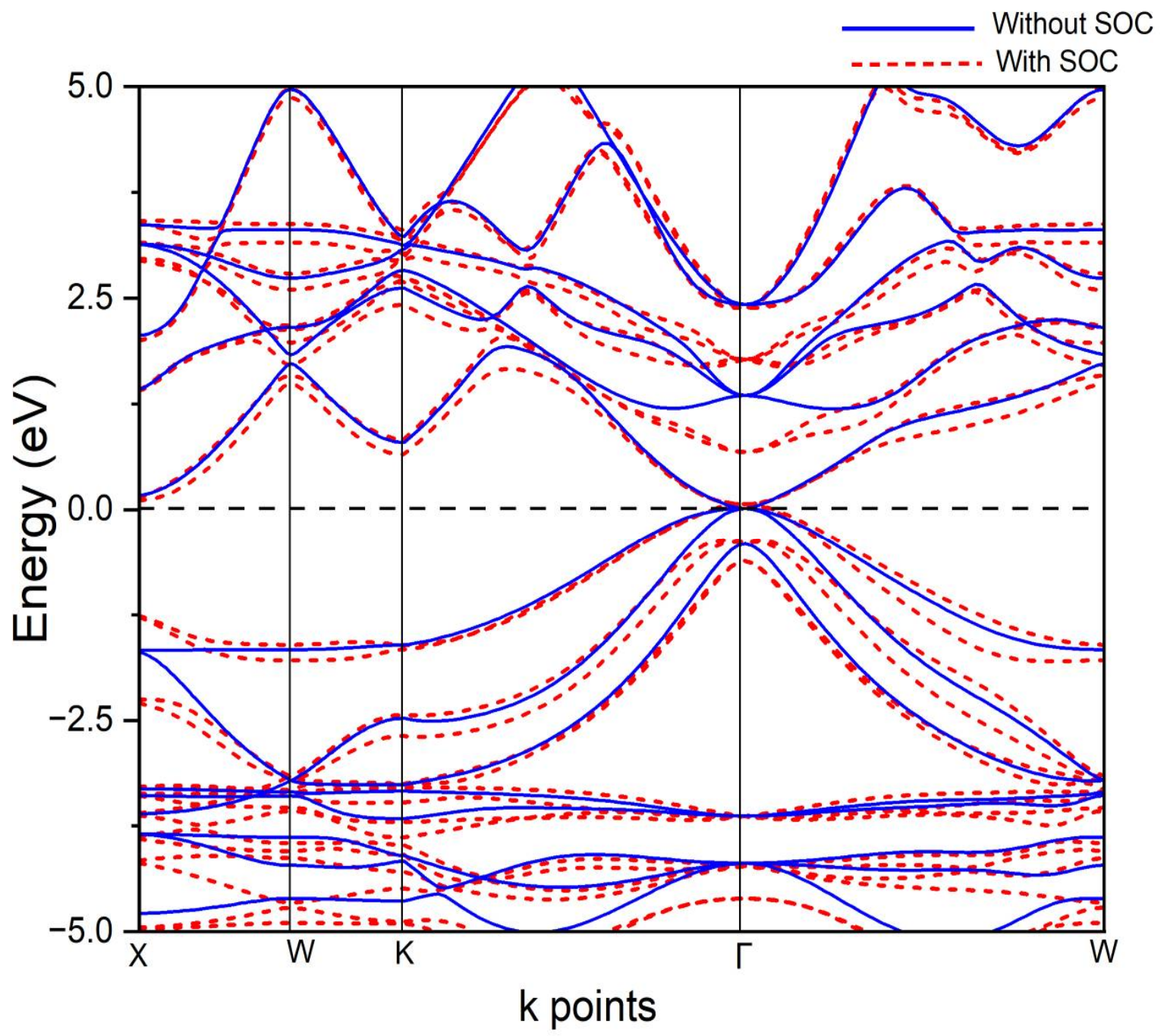

**Figure 3:** The electronic band structure of HoPdBi compound with and without SOC (obtained using QE). The dashed horizontal black line marks the Fermi energy (set to 0 eV).

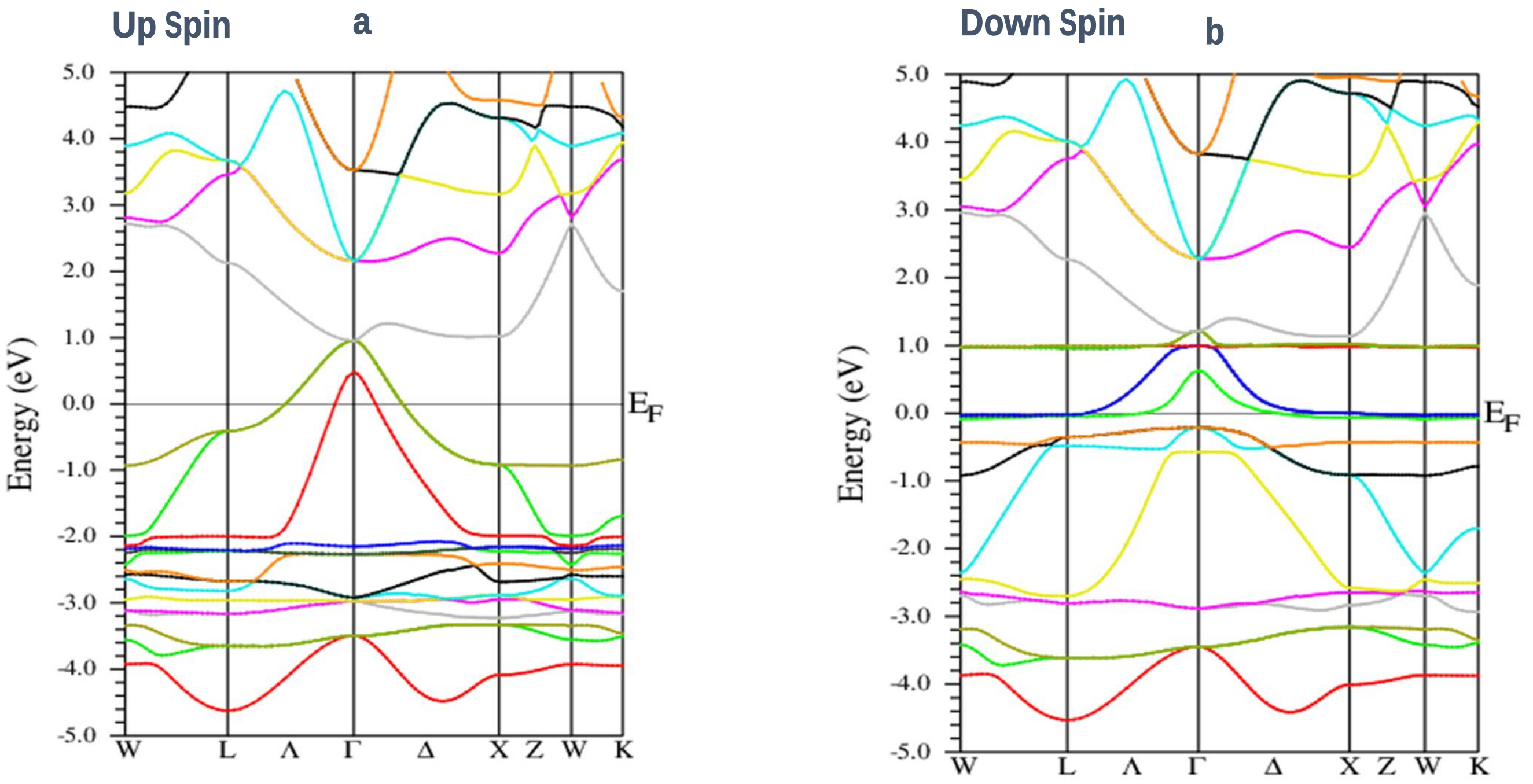

**Figure 4:** The electronic band structures of HoPdBi (a) for spin up and (b) for spin down (obtained using Wien2k) configurations. The solid horizontal line marks the Fermi energy (set to 0 eV).

From Fig. 3, it is seen that SOC has significant effect on the band dispersions. SOC lifts band degeneracy significantly. SOC led to strong shifts in energy for certain branches in the valence band and conduction band. But close to the Fermi level, the effect of SOC is minimal. In Fig.4(a, b) we see the effect of spin polarized calculations. For up spin and down spin, the compound remains metallic but with different degree. The energy dispersion curves are also different for the two spin configurations close to the Fermi energy. The asymmetry in the band structure should lead to magnetic order in HoPdBi.

A material's total density of states (TDOS) and partial electronic density of states (PDOS) must be known in order to properly understand the bonding nature of atoms and the role that electrons play in electronic conductivity, thermal conductivity, and optical characteristics. Additionally, the DOS establishes a number of physical variables that are directly related to the electronic density of states at the Fermi level, such as the electronic contribution to a metal's heat capacity, spin paramagnetic susceptibility, and possible magnetic order [53,55]. The HoPdBi compound's TDOS and PDOS are displayed in Figure 5. The Fermi energy is shown by the vertical dashed line at zero energy level.

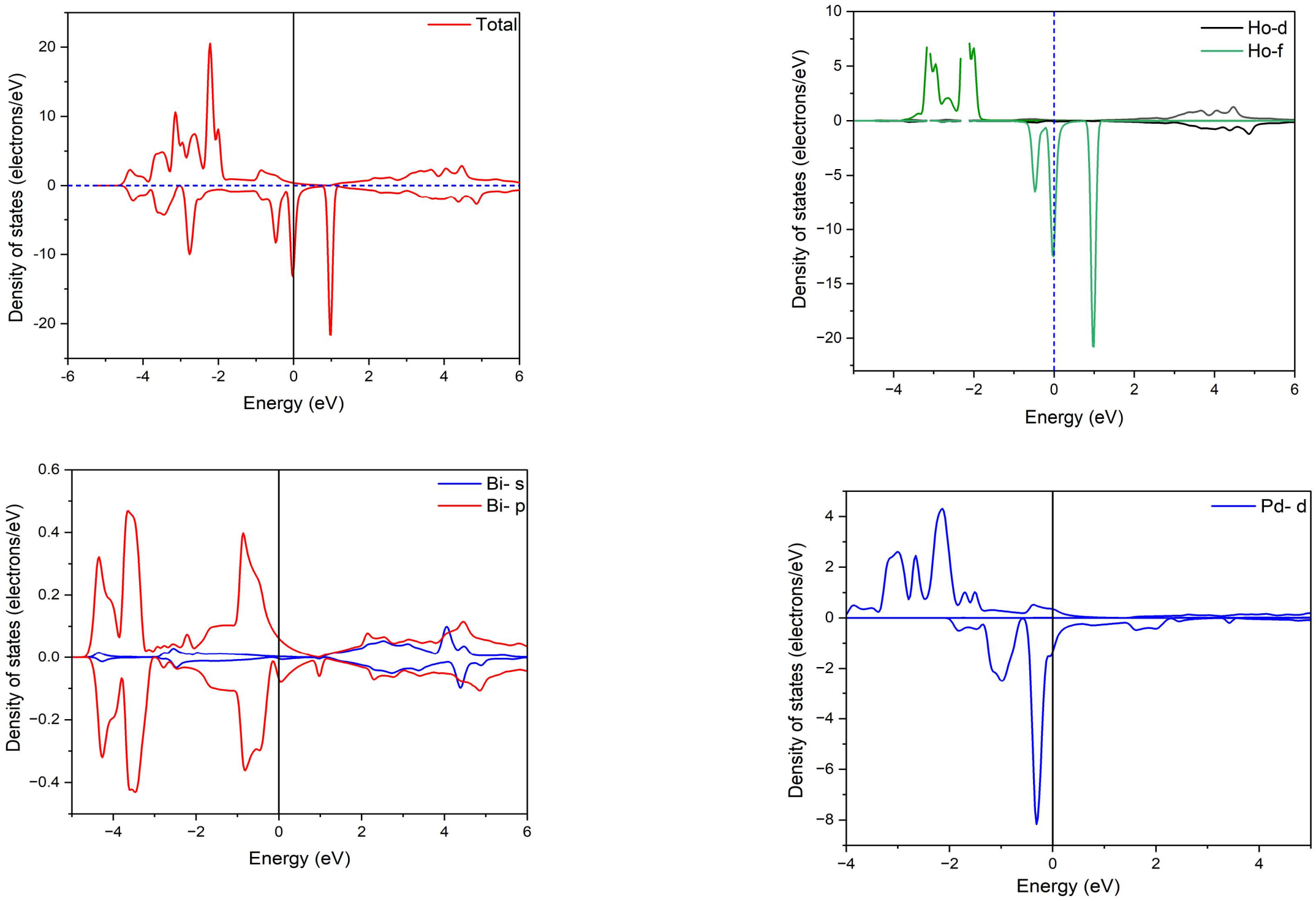

**Figure 5:** Total density of states (TDOS) and partial density of states (PDOS) of HoPdBi (obtained using Wien2K) as a function of energy for both up and down spins. Fermi level is set to 0 eV.

For up and down spin, there is significant difference in density of states at Fermi level. This indicates that the compound is suitable for spintronic applications. Ho-f orbital contributes the maximum to the TDOS whereas Bi-p orbital contributes minimum. The difference in the electronic density of states between the spin up and spin down orbitals at $E_F$, suggests the presence of spin polarization. The following relation can be used to compute the energy-dependent degree of electronic spin polarization [56].

$$P_e(E) = \frac{|n_\uparrow(E) - n_\downarrow(E)|}{n_\uparrow(E) + n_\downarrow(E)} \times 100\% \qquad (20)$$

where the density of states for up and down spin states is denoted by $n_\uparrow(E)$ and $n_\downarrow(E)$, respectively. At the Fermi level, electron spin polarization for HoPdBi is 93.77%. Such high spin-polarization is recognized as a selection criterion of materials for spintronic applications.

### 3.4 Magnetic properties

In the present study, the magnetic response [56] of HoPdBi compound was predicted using spin-polarized calculations for the magnetic moment in Bohr magneton ($\mu_B$). Table 5 lists the computed interstitial, local, and total magnetic moments of HoPdBi compound. The total magnetic moments $\mu^{total}$ is slightly influenced by the interstitial magnetic moments $\mu^{int}$ for each of the elements. According to Hund's and Aufbau's principle, the growing quantity of unpaired electrons from Ho may be the cause of the HoPdBi bulk magnetic moments [57]. There is a relationship between the magnetic moment and DOS [58]. The f-state of Ho is responsible for the majority of the TDOS at the Fermi level, with the d-state of Pd and the p-state of Bi contributing only slightly. Rare-earth (RE) and other atomic orbitals of HoPdBi are hybridized as a result of several contributions. Because coulomb interaction may raise the local magnetic moment and the localization of the corresponding f-orbitals at RE sites, thereby decreasing the magnetic moments at Pd and Bi sites; we conclude that the magnetic moment value is larger for the rare earth element.

**Table 5** Calculated total magnetic moment ($\mu^{total}$), interstitial ($\mu^{int}$), and magnetic moments of Ho, Pd, and Bi atoms (in Bohr magneton $\mu_B$) of the HoPdBi compound.

| Compound | Scheme | $\mu^{int}$ | $\mu^{Ho}$ | $\mu^{Pd}$ | $\mu^{Bi}$ | $\mu^{total}$ |
|---|---|---|---|---|---|---|
| HoPdBi | GGA-PBE | 0.0649 | 3.6265 | 0.0165 | 0.0047 | 3.7127 |

### 3.5 Fermi surface

The Fermi surface HoPdBi is shown in Fig. 6. All the bands crossing the Fermi level contribute to the Fermi surface. The overall Fermi surface topology is complex, both electron- and hole-like Fermi sheets are observed.

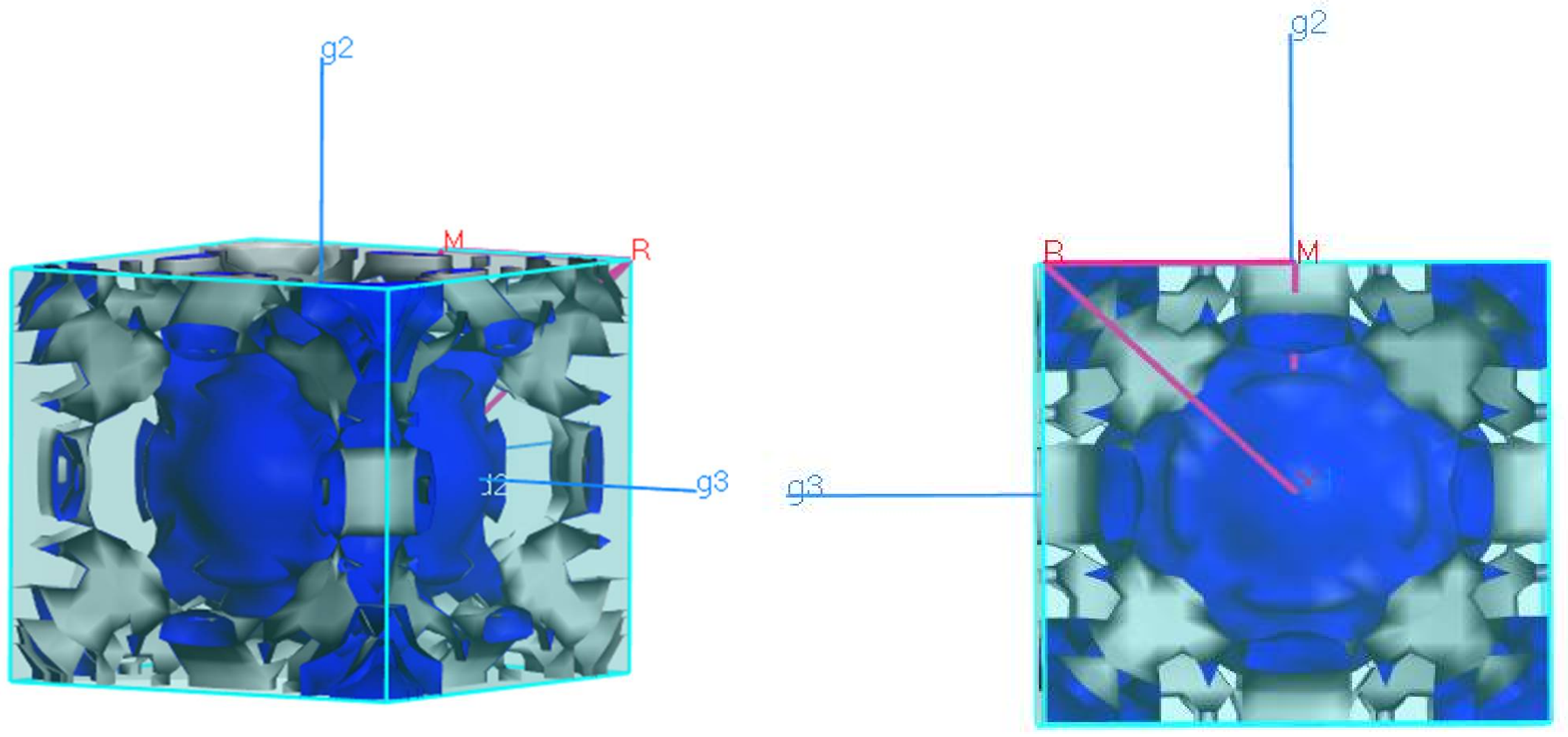

**Figure 6:** Fermi Surface of HoPdBi. The symmetry directions in the Brillouin zone are shown.

The symmetry directions in the Brillouin zone are marked, showing how the Fermi surface evolves across the Brillouin zone.

### 3.6 Phonon dynamics and acoustic properties

#### 3.6.1 Phonon dispersion curves and phonon density of states

Understanding the physical properties of crystalline materials requires an understanding of phonon characteristics. The energy quanta of lattice vibrations are called phonons. The phonon spectrum affects electrical, thermal, elastic, and lattice dynamical properties. The phonon spectrum is determined by the crystal symmetry, stiffness constants and mass of the constituent atoms [49,59,60]. The electron–phonon interaction function is directly related to the phonon density of states (PHDOS). From phonon dispersion spectra (PDS) we can determine the dynamical stability of a crystal structure. It gives indication of possible structural phase transition and thermal properties of a solid. By using the linear perturbative approach [61], we have calculated the phonon dispersion spectra and phonon density of states of HoPdBi compound along the high symmetry directions of the first Brillouin zone. The results are shown in Figs. 7(a) and 7(b). For HoPdBi, the structure is dynamically stable since every phonon mode in the first BZ is positive. The slope of the acoustic dispersion relation, $\partial\omega/\partial k$, provides the speed of propagation of an acoustic phonon, which is also the sound speed in the lattice. The dispersion relation is nearly linear at low values of $k$ (i.e., in the long wavelength limit), regardless of the phonon frequency. This behavior does not hold at large values of $k$, i.e., for short wavelengths. These short wavelength phonon modes are generally the optical modes. The lower branches in phonon dispersion spectra are the acoustic branches and the upper branches represent the optical branches. The acoustic branches are designed to allow the lattice's atoms to move in phase with respect to their equilibrium positions. The frequency of the acoustic modes at G-point is zero. This is another indication that the compound under study is dynamically stable. The optical branches play a major role in controlling the optical properties of crystals [62]. The number of atoms present in HoPdBi compound is 3 so there are 3 acoustic modes and 6 optical modes present. There is no phonon gap in Fig. 7(a). The high PHDOS (Fig. 7(b)) regions for the acoustic branches at low phonon frequencies assist significantly in the thermal transport. The vibration of light Pd atoms in the crystal is responsible for the highest energy phonon dispersion branches of HoPdBi. In the optical region, the PHDOS is found to be rather high at frequency of 4.15 THz (Fig. 7(b)). It is anticipated that these phonons will be crucial in defining HoPdBi's optical characteristics.

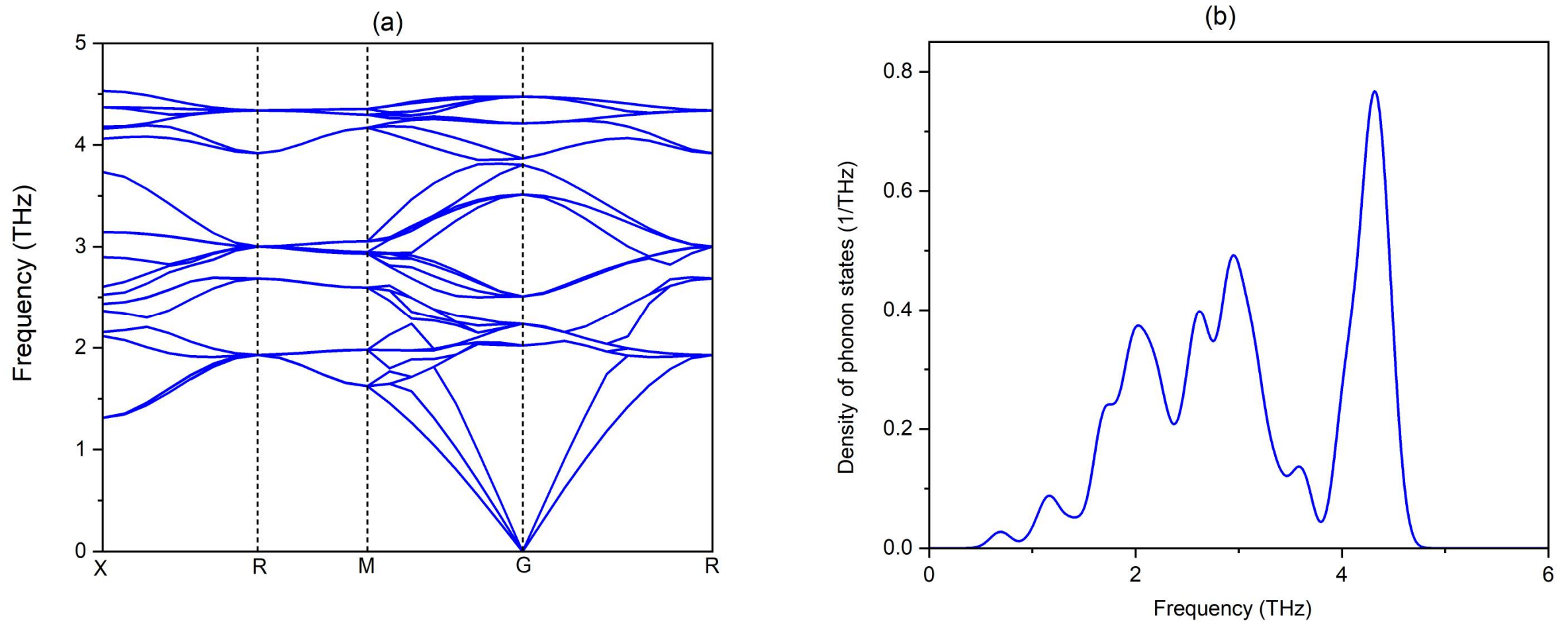

**Figure 7:** Calculated (a) phonon dispersion spectra and the (b) PHDOS for HoPdBi compound at zero pressure.

### 3.6.2. Acoustic properties

The sound velocity through a material is an important parameter to understand the thermal and acoustic behaviors. The average sound velocity in solids, $V_m$, is related to the shear modulus, bulk modulus and crystal density. The $V_m$ is given by the harmonic mean of the average longitudinal and transverse sound velocities, $V_l$ and $V_t$. The relevant relations are given below [63]:

$$V_m = [\frac{1}{3}(\frac{1}{V_l^3} + \frac{2}{V_t^3})]^{-1/3} \quad (21)$$

$$V_l = [\frac{3B+4G}{3\rho}]^{1/2} \quad (22)$$

$$V_t = [\frac{G}{\rho}]^{1/2} \quad (23)$$

Table 6 exhibits the calculated crystal density of and the acoustic velocities in HoPdBi.

**Table 6:** Density ρ ($Kg/m^3$ ), transverse velocity $V_t$ ($ms^{-1}$ ), longitudinal velocity $V_l$ ($ms^{-1}$ ) and average elastic wave velocity $V_m$ ($ms^{-1}$) of HoPdBi.

| Compound | $\rho$ | $V_l$ | $V_t$ | $V_m$ | Ref. |
|---|---|---|---|---|---|
| HoPdBi | 10800 | 3309.77 | 17011.07 | 1913.46 | This work |

### 3.7 Optical properties

Finding possible optoelectronic applications requires a thorough understanding of a system's sensitivity to visible, ultraviolet, and infrared spectra [64-68]. This understanding is gained through the study of energy/frequency dependent optical parameters. These parameters are largely controlled by the electronic band structure. In each optical calculation, we have employed 0.5 eV Gaussian smearing, along with an empirical Drude damping of 0.05 eV. The damping term is necessary to include metallic contributions.

The complex dielectric function ε(ω), refractive index n(ω), optical conductivity σ(ω), reflectivity R(ω), absorption coefficient α(ω), and energy loss function L(ω) are the frequency/energy dependent optical constants that are explored in this section. The optical characteristics of the compound of interest are displayed in Figure 8 for incident energy up to 30 eV. Owing to its cubic structure HoPdBi is optically isotropic.

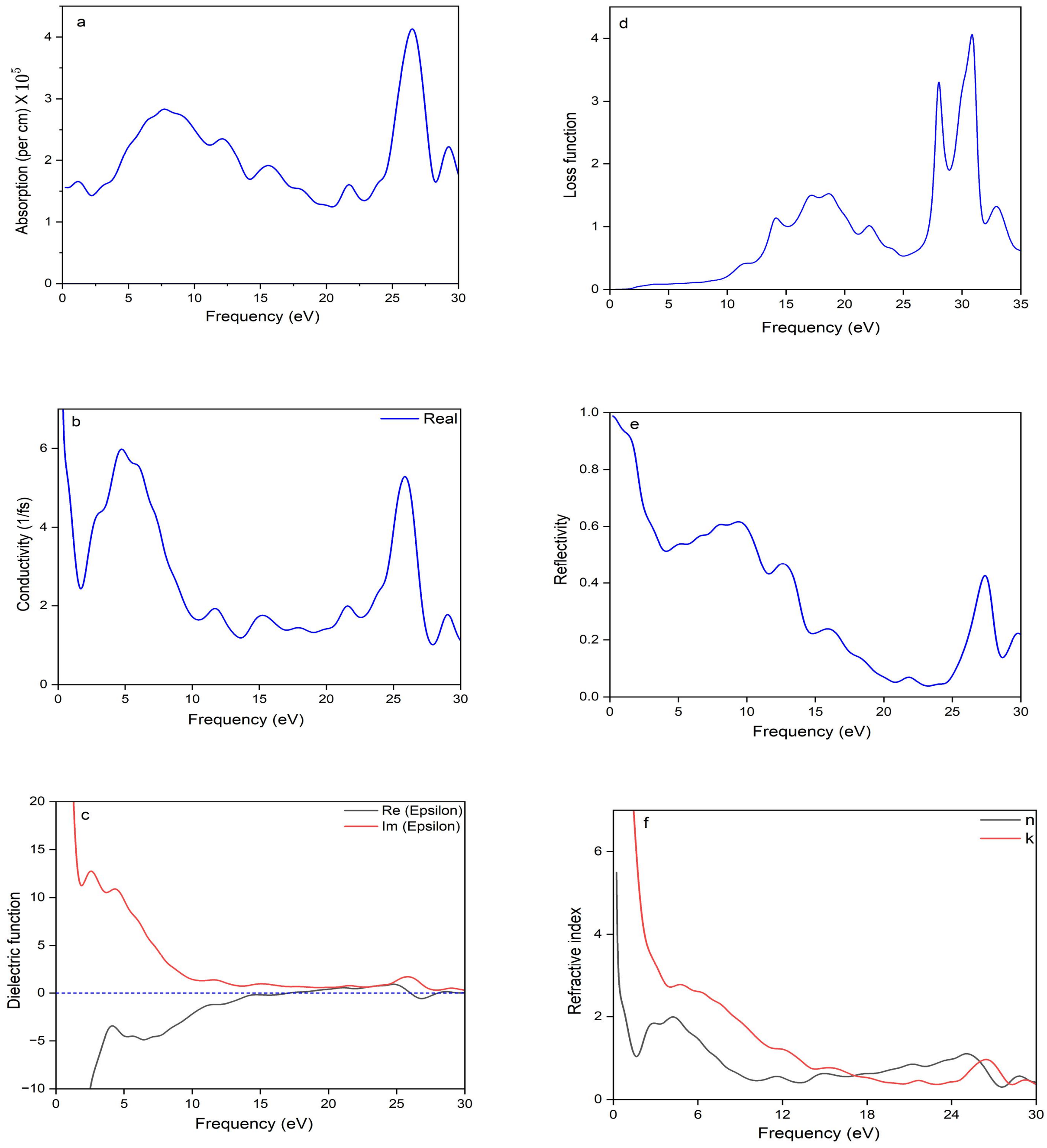

The amount of light with a given energy (wavelength/frequency) that can enter a material before being absorbed is determined by the absorption coefficient α(ω). Figure 8(a) displays the HoPdBi compound's absorption coefficient. Photon absorption is found to begin at zero energy. This suggests that HoPdBi does not have an optical bandgap. This further demonstrates that it is a metal. In the ultraviolet range between 5 eV and 13 eV, we also observe a noticeable absorption implying that the compound is an efficient absorber of ultraviolet light.

The optical conductivity of charge carriers characterizes their dynamic response to incident light. The real part of optical conductivity σ(ω) are depicted in Figure 8(b). The optical conductivity of HoPdBi compound also starts at zero photon energy, demonstrating the absence of a band gap, which is consistent with the electronic band structure and TDOS calculations and supports the metallic nature of the material under investigation. The optical conductivity spectrum is characterized by two peaks at ~5 eV and ~25 eV. The optical conductivity of HoPdBi is very high in the infrared region.

The real and imaginary parts of the dielectric constant are depicted in Figure 8(c). While the real part of the dielectric function is linked to anomalous dispersion and electrical polarization, the imaginary part is related to the electromagnetic wave's energy dissipation within the medium. The metallic nature of the HoPdBi compound is suggested by the negative value of the real component at low energies. We may predict that the energy corresponding to the plasma frequency of HoPdBi is around 30 eV since both the real and imaginary parts approach zero at that energy. Consequently, HoPdBi becomes relatively transparent for photon energies greater than 30 eV.

The loss functions L(ω) of the HoPdBi compound is displayed in Figure 8(d). A fast electron moving through a material can excite collective charge oscillation modes. This results in energy loss of the electron characterized by the peaks in the loss function L(ω) [77,69]. There are sharp loss peaks between 27 eV to 31 eV. The plasmon energy is represented by the locations of these loss peaks. The collective oscillation of electrons in the jellium model causes plasma oscillations at these specific energies. It is noteworthy that abrupt decreases in the reflectance and absorption coefficient occur in tandem with the plasma resonance energies.

The amount of electromagnetic wave (EMW) energy reflected from a surface compared to the energy it had when incident on the surface is known as the reflectivity. Figure 8(e) illustrates the reflectivity profiles for HoPdBi compound. At 0 eV, the maximum reflectivity value in HoPdBi is around ~98%. From infrared to near-ultraviolet (up to ~2.11 eV), the reflectivity spectra stay over ~75%. So, HoPdBi compound can be used as a very good solar radiation reflector.

The real and imaginary components of the HoPdBi compound's refractive index are displayed in Figure 8(f). The phase velocity of an EMW is controlled by the real part of the refractive index, whereas the attenuation of the EMW within the material is controlled by the imaginary part, also referred to as the extinction coefficient. At low energies, the HoPdBi compound shows a high value of the real component of the refractive index, which is in the infrared regions. This is noteworthy. Applications for optoelectronic devices benefit from such high values [70].

### 3.8 Thermo-physical properties

#### (a) Debye temperature

A solid's bonding forces, energy of vacancy formation, melting point, thermal conductivity, phonon dynamics, specific heat, and superconductivity, among other thermo-physical characteristics, depends on its Debye temperature ($\Theta_D$). The temperature at which the phonon wavelengths of a solid roughly match the average interatomic distance is known as the Debye temperature. Lattice dynamics can be divided into high- and low-temperature zones using this temperature. The Debye temperature additionally differentiates between the lattice vibration's classical and quantum characteristics. When T is greater than $\Theta_D$, the energy of all vibrational modes is approximately $K_B T$. At T < $\Theta_D$ however, the higher frequency modes are not excited. There are several methods for calculating $\Theta_D$. Debye temperature from specific heat measurement and the Debye temperature determined from the elastic constants coincide when low temperature vibrational excitations caused only by acoustic modes are present. Here, we have determined the Debye temperature of HoPdBi using the following expression [71]:

$$\Theta_D = \frac{h}{K_B}[(\frac{3n}{4\pi})\frac{N_A\rho}{M}]^{1/3}V_m \quad (24)$$

where h is the Planck’s constant, $K_B$ is the Boltzmann’s constant, n denotes the number of atoms within the unit cell, M is molar mass, ρ is density, $N_A$ is Avogadro’s number and $V_m$ denotes mean sound velocity.

In previous section we already have determined the value of $V_m$, $V_l$ and $V_t$. So, using these values we can determine the Debye temperature $\Theta_D$ which is 195.16 K for HoPdBi. This value is quite low, consistent with soft nature of the compound under study.

**(b) The melting temperature**

One of the important characteristics that determines the temperature range over which a solid can be used is the melting temperature ($T_m$). A solid with a high melting temperature has low coefficient of thermal expansion, high cohesive energy, and a high bonding energy [72]. The bonding strength has a direct bearing on it. Solids can be utilized continuously at temperatures lower than $T_m$ without experiencing oxidation, chemical change, or severe distortion that could lead to mechanical failure. The following formula [72] can be used to determine the melting temperature $T_m$ of solids using the elastic constants:

$$T_m = \left[553K + \left(5.91\frac{K}{GPa}\right)C_{11}\right] \qquad (25)$$

The computed value of the melting temperature is given in Table 7.

**(c) Lattice thermal conductivity**

In materials, phonons and electrons can both carry thermal energy. In metals, electrons are the primary heat carriers at low temperatures. When temperatures are high, the lattice contribution becomes more significant. For high-temperature applications, determining a material's lattice thermal conductivity is essential. The lattice thermal conductivity ($K_{ph}$) of a material determines how much heat energy is transferred by lattice vibration when a temperature gradient is present. Many engineering specialties use materials with both low and high lattice thermal conductivity, depending on the nature of application. Recently, there has been interest in finding low thermal conducting materials to enhance the performance of thermal barrier coatings (TBC), thermoelectric devices, and solid-state refrigeration. However, high thermal conductivity materials with little heat waste are needed to improve the efficiency of heat removal in microelectronic and nano electronic devices. The $K_{ph}$ as a function of temperature can be estimated using the following formula, which was developed by Slack [73]:

$$K_{ph}(T) = A(\gamma)\frac{M_{av}\Theta_D{}^3\delta}{\gamma^2 n^{\frac{2}{3}} T} \quad (26)$$

In this equation, $M_{av}$ is the average atomic mass in kg/mol, $\Theta_D$ is the Debye temperature in K, δ is the cubic root of average atomic volume in meter, n refers to the number of atoms in the conventional unit cell, T is the absolute temperature in K, and $\gamma$ is the acoustic Grüneisen parameter which determines the degree of anharmonicity of phonons. Grüneisen parameter is a crucial parameter in lattice dynamics and thermodynamics due to its relationships with the volume, bulk modulus, heat capacity, and thermal expansion coefficient. Low phonon anharmonicity results in good thermal conductivity. The Grüneisen parameter which is a dimensionless quantity and, can be derived from the Poisson's ratio using the equation [74] given below:

$$\gamma = \frac{3(1+\nu)}{2(2-3\nu)} \quad (27)$$

with

$$A(\gamma) = \frac{5.720\times10^5\times0.849}{2\times(1-\frac{0.514}{\gamma}+\frac{0.228}{\gamma^2})} \quad (28)$$

Calculated room temperature (300 K) lattice thermal conductivity and the Grüneisen parameter are listed in Table 7.

**(d) Minimum thermal conductivity**

The lowest thermal conductivity is a limit of a basic thermal property. The minimum thermal conductivity ($K_{min}$), a minimal value for a compound's thermal conductivity, is attained at high temperatures exceeding the Debye temperature. It is important to note that crystal defects (such dislocations, individual vacancies, and long-range strain fields related to impurity inclusions and dislocations) have no effect on the minimal thermal conductivity [75 – 77]. Clarke used the Debye model of compounds at high temperatures to derive the following formula to get the lowest thermal conductivity ($K_{min}$):

$$K_{min} = K_B \nu_m \,(V_{atomic})^{-2/3} \quad (29)$$

In this equation, $K_B$ is the Boltzmann constant, $\nu_m$ is the average sound velocity and $V_{atomic}$ is the cell volume per atom.

**(e) Dominant phonon wavelength**

The wavelength at which the phonon distribution function reaches its maximum is known as the dominant phonon's wavelength ($\lambda_{dom}$). Low crystal density, high shear modulus, and high mean sound velocity lead to large values of the dominating phonon wavelength. We have used the following relationship [35,78] to compute $\lambda_{dom}$ for HoPdBi at 300 K:

$$\lambda_{dom} = \frac{12.566 v_m}{T} \times 10^{-12} \quad (30)$$

where, $v_m$ is the average sound velocity in $ms^{-1}$, T is the temperature in K. The estimated values of $\lambda_{dom}$ in meters and minimum thermal conductivity are also listed in Table 7.

**Table 7**: Calculated melting temperature $T_m$ (K), Debye temp. $\Theta_D(T)$, $K_{ph}(W/m.K)$, $K_{min}(W/m.K)$ and wavelength of the dominant phonon mode $\lambda_{dom}$ (m) at 300 K of HoPdBi compound.

| Compound | $T_m$ | $\Theta_D$ | $K_{ph}$ | $K_{min}$ | $\lambda_{dom}$ | Ref. |
|---|---|---|---|---|---|---|
| HoPdBi | 1253 | 195.16 | 4.64 | 0.312 | $8.01 \times 10^{-11}$ | This work |

It is seen from Table 7 that the melting temperature is moderate. The minimum thermal conductivity is very low; suitable for a TBC material [79-81]. Therefore, HoPdBi has potential to be used as TBC below ~1200 K.

## 4. Conclusions

This research presents a detailed first-principles investigation of the half-Heusler HoPdBi compound employing the density functional theory. The results reported here are mostly novel. In cases where they were accessible, we compared our findings with previous results and fair agreements were found. The HoPdBi compound is elastically stable and shows isotropic nature. The HoPdBi compound is soft, ductile and machinable. The value of $\mu_M$ in HoPdBi compound implies a good level of dry lubricity. Ductility, machinability, and hardness values of this compound suggest that this compound is suitable for engineering applications. Clear semi-metallic features are suggested by the electronic band structure and the electronic energy density of states. High degree of spin polarization has been found. The magnetic moments associated with the elements have been calculated. Due to spin-asymmetric electronic energy density of states HoPdBi is suitable for spintronic applications. The electronic band structure and DOS profiles are in excellent agreement with the optical constant profiles of HoPdBi. Infrared-visible light is efficiently-reflected by HoPdBi. The compound absorbs UV light efficiently. At low photon energies, the refractive index is high. Applications for optoelectronic devices benefit from such high values [70]. Multiple loss peaks imply multiband contributions to the plasma resonance.

In conclusion, we have offered numerous new findings about the different characteristics of the HoPdBi compound. Our findings should encourage materials scientists to further explore this fascinating system through both theoretical and experimental means.

**Acknowledgements**

S.H.N. acknowledges the research grant (1151/5/52/RU/Science-07/19-20) from the Faculty of Science, University of Rajshahi, Bangladesh, which partly supported this work. This work is dedicated to the cherished memory of the martyrs of the July-August 2024 revolution in Bangladesh, whose sacrifices will forever inspire us.

**Conflict of interest statement**

The authors declare no competing interests.

**Credit author statement**

**Tanvir Khan**: Methodology, Software, Writing- Original draft. **F. Parvin**: Methodology, Supervision, Writing- Reviewing and Editing. **S. H. Naqib**: Conceptualization, Supervision, Formal Analysis, Writing- Reviewing and Editing.

**Data availability statement**

The data sets generated and/or analyzed in this study are available from the corresponding author upon reasonable request.